\documentclass{article}

\usepackage{setspace}
\usepackage[english]{babel}
\usepackage[utf8]{inputenc}
\usepackage[T1]{fontenc}
\usepackage{multimedia}
\usepackage{amsfonts,multirow}
\usepackage{amsmath,bm}
\usepackage{amsthm}
\usepackage{amssymb}
\usepackage{epsfig}
\usepackage{aeguill}
\usepackage{graphicx, psfrag, color}
\usepackage{subfig}

\usepackage{amsbsy,fancybox}
\usepackage{color}

\usepackage{geometry}

\definecolor{bordeaux}{rgb}{0.25,0.15,0.75}
\definecolor{bluegray}{rgb}{0.25,0.15,0.75}
\definecolor{greengray}{rgb}{0.05,0.50,0.15}
\definecolor{darkbrown}{rgb}{0.35,0.05,0.05}
\definecolor{puorange}{rgb}{0.80,0.20,0}
\definecolor{black}{rgb}{0,0,0}

\newcommand{\cL}{{\cal L}}

\newcommand{\bu}{{\bf u}}

\newcommand{\bbeta}{{\boldsymbol \beta}}
\newcommand{\bdelta}{{\boldsymbol \delta}}
\newcommand{\bDelta}{{\boldsymbol \Delta}}
\newcommand{\bSigma}{{\boldsymbol \Sigma}}
\newcommand{\bgamma}{{\boldsymbol \gamma}}
\newcommand{\bGamma}{{\boldsymbol \Gamma}}

\newcommand{\bdeltak}{\bdelta_{k}}

\newcommand{\hbbeta}{\widehat {\boldsymbol \beta}}
\newcommand{\hbeta}{\widehat {\beta}}
\newcommand{\hdelta}{\widehat {\delta}}

\newcommand{\hbgamma}{\widehat {\boldsymbol \gamma}}
\newcommand{\hbdelta}{\widehat {\boldsymbol \delta}}

\newcommand{\bnu}{\boldsymbol \nu}

\newcommand{\hmu}{\widehat \mu}
\newcommand{\bmu}{\boldsymbol \mu}
\newcommand{\hbmu}{\widehat {\bmu}}

\newcommand{\btau}{{\boldsymbol \tau}}

\newcommand{\bX}{{\bf X}}
\newcommand{\bx}{{\bf x}}
\newcommand{\bW}{{\bf W}}

\newcommand{\LL}{{\mathcal L}}

\newcommand{\bcX}{{\boldsymbol {\mathcal X}}}

\renewcommand{\P}{{\rm I}\kern-0.12em{\rm P}}
\newcommand{\1}{{\rm 1}\kern-0.28em{\rm I}}
\newcommand{\E}{{\rm I}\kern-0.12em{\rm E}}
\newcommand{\R}{{\rm I}\kern-0.14em{\rm R}}
\newcommand{\N}{{\rm I}\kern-0.14em{\rm N}}
\newcommand{\reals}{{\rm I}\kern-0.16em{\rm R}}

\begin{document}

\begin{center}
{\huge {\bf Sparse estimation for case-control studies with multiple subtypes of cases.}}
\vskip10pt
{\Large Nadim Ballout, Cedric Garcia and Vivian Viallon\footnote{Corresponding Author: viallonv@iarc.fr}}
\end{center}

\begin{abstract}
The analysis of case-control studies with several subtypes of cases is increasingly common, e.g. in cancer epidemiology. For matched designs, we show that a natural strategy is based on a stratified conditional logistic regression model. Then, to account for the potential homogeneity among the subtypes of cases, we adapt the ideas of data shared lasso, which has been recently proposed for the estimation of regression models in a stratified setting. For unmatched designs, we compare two standard methods based on L1-norm penalized multinomial logistic regression. We describe formal connections between these two approaches, from which practical guidance can be derived. We show that one of these approaches, which is based on a symmetric formulation of the multinomial logistic regression model, actually reduces to a data shared lasso version of the other. Consequently, the relative performance of the two approaches critically depends on the level of homogeneity that exists among the subtypes of cases: more precisely, when homogeneity is moderate to high, the non-symmetric formulation with controls as the reference is not recommended. Empirical results obtained from synthetic data are presented, which confirm the benefit of properly accounting for potential homogeneity under both matched and unmatched designs. We also present preliminary results from the analysis a case-control study nested within the EPIC cohort, where the objective is to identify metabolites associated with the occurrence of subtypes of breast cancer.
\end{abstract}

\section{Introduction}
\label{s:intro}

The rise of -omics and other high-dimensional data (image, reimbursement claims, etc.) in medical science gives researchers access to numerous features that may predict outcomes of interest, like cancer development. However, this relatively cheap source of information comes at a price: the curse of dimensionality makes multivariate modeling of such data impossible without further assumptions. In other words, some prior information has to be properly accounted for to reduce dimensionality and accurately estimate high-dimensional multivariate models. Under parametric regression models, one common prior information, or assumption, is sparsity of the parameter vector. The use of  $L_1$-norm regularized approaches has been shown to yield optimal estimates when the true vector is sparse, under technical assumptions on the design matrix \cite{Wainwright2009sharp,bach2010self,BRT2009}. As a result, $L_1$ penalized logistic models \cite{mccullagh1989generalized,ParkHastie,wu2009genome} are now standard tools when studying risk factors of a disease in a high-dimensional setting.  

However, for many diseases that were primarily considered as one single disease (breast cancer, colorectal cancer), several subtypes have now been recognized. They can either be histological, as for breast cancer, or anatomical, as for colorectal cancer. Even if commonalities may exist among these subtypes, they have their own specificities regarding both prognosis and etiology. For example, the cancer epidemiology community is now increasingly concerned with the identification of subtype specific risk factors for various cancer sites. This is the case in our motivating example presented in Section \ref{sec:App}, which deals with the identification of metabolites associated with breast cancer subtypes, based on a matched case-control study nested in the EPIC (European Prospective Investigation into Cancer and nutrition) cohort study. 

For unmatched case-control studies with multiple subtypes of cases, a natural extension of the binary logistic regression model is the multinomial logistic regression model \cite{mccullagh1989generalized,BecgGray}. If $K-1$ denotes the number of subtypes for a given integer $K>1$, inference under this model consists in estimating $K-1$ parameter vectors of size $p$, where $p$ denotes the number of covariates (which may include interactions as well as an intercept term). On the other hand, for matched case-control studies with $K-1$ subtypes, the total sample can be decomposed into $K-1$ sub-samples, one for each subtype.  Assuming for simplicity a 1:1 matching design, each sub-sample is made of pairs composed by one case of one particular subtype and one matched control. Then, each sub-sample can be analyzed separately, e.g. by applying a sparse conditional logistic regression model \cite{Avalos}. Again, the overall analysis boils down to the estimation of $K-1$ parameter vectors of size $p$. 

Because commonalities exist between the subtypes of cases, some level of homogeneity is expected among those $K-1$ parameter vectors, in both the matched and unmatched settings. Properly accounting for this homogeneity is key to reduce the dimensionality and improve estimation efficiency. In the matched setting, $K-1$ sparse conditional logistic regression models have to be estimated on $K-1$ sub-groups, where these sub-groups are defined according to the subtype of the case of each pair (see Section \ref{sec:Matched} for more details). Then, inference falls into the framework of stratified regression modeling for which data shared lasso has recently been developed as a way to account for the expected homogeneity among the $K-1$ parameter vectors to be estimated \cite{Gross,AutoRefLasso}. Under linear models, data shared lasso has been shown to enjoy good theoretical and empirical properties \cite{AutoRefLasso}. In addition, data shared lasso is easy to implement because it can be rewritten as a standard lasso after a simple transformation of the original data. 

In this article, we will show how the ideas of data shared lasso can be applied to analyze case-control studies when multiple subtypes of cases are present. In Section \ref{sec:Matched}, we start with the matched design and we show how data shared lasso can be used to estimate stratified sparse conditional logistic models. Section \ref{sec:UnMatched} is devoted to the unmatched setting, under which sparse multinomial logistic regression models are natural, as mentioned above. Actually, two formulations of sparse multinomial logistic regression models have been proposed in the literature. A first one, which we will refer to as the standard one, relies on the selection of a reference category and the estimation of $(K-1)$ parameter vectors \cite{BecgGray}. Alternatively, a more symmetric formulation of the model can be adopted, where no reference category has to be selected and $K$ parameter vectors are to be estimated \cite{glmnet}. Unpenalized estimation is impossible under this over-parametrized model due to a clear lack of identifiability. However, $L_1$-penalized estimation can be performed, as implemented in the popular glmnet R package. To our best knowledge, no clear guidance exists in the literature on how to chose between the two formulations of sparse multinomial logistic regression models. We will formally establish that the $L_1$-penalized strategies associated with the two formulations differ in the way they account for potential homogeneity among the parameter vectors to be estimated. More precisely, we show in Section \ref{sec:UnMatched} that  $L_1$-penalized estimates derived under the symmetric formulation coincide with the estimates derived under the standard formulation when using a data shared lasso penalty. In Section \ref{sec:Simul}, we present results from a simulation study, which illustrate the interest of data shared lasso estimates when homogeneity exists among the parameter vectors to be estimated, under both the matched and unmatched settings. Section \ref{sec:App} is devoted to our illustrative example. Finally, concluding remarks are provided in Section \ref{sec:Discussion}.

\section{Matched case-control studies with multiple subtypes of cases and stratified conditional logistic models}\label{sec:Matched}

Conditional logistic regression is a standard tool for the analysis of matched case-control studies when a single type of cases is considered \cite{PearceBMJ,ModernEpidemiology}. Here, we show how the ideas of data shared lasso can be applied to handle the situation where $K-1$ subtypes of cases are present, for some given integer $K>1$. 

\subsection{Setting}

Consider a matched case-control study where information about subtype is available for each case. We denote the number of subtypes by $K-1$, for some given integer $K>1$, and we will use the notation $[I]=\{1, \ldots, I\}$ for any integer $I\geq 1$. For simplicity, we further assume a 1:1 matched case-control design where matching is based on some variables $\bW\in\R^q$. Denoting by $n\geq 1$ the total sample size, the sample then consists of $m=n/2$ pairs of individuals. In this matched setting with $K-1$ subtypes of cases, the total sample can be divided into $K-1$ sub-samples. For any $k\in [K]$, the $k$-th sub-sample ${\cal M}_k$ is made of the $m_k$ pairs composed by each case of subtype $k$ and his matched control.  These sub-samples naturally define sub-groups, or strata, in the total sample. We should however stress that these strata differ from the ``usual''  strata defined in the context of conditional logistic regression models, the latter corresponding to case-control pairs. In other respect, for future use, we introduce the categorical variable $S$ which takes values in $[K-1]$ and indicates the sub-sample to which an observation belongs. In other words, $S=k$ for all observations in ${\cal M}_k$.


Let us first focus on the $k$-th stratum ${\cal M}_k$, which is made of cases of subtype $k$ and their matched controls. For any matched pair $\ell$ of observations belonging to ${\cal M}_k$, denote by $\bx^{(k)}_{\ell,i}\in\R^p$, for some $p\geq1$, and $Y^{(k)}_{\ell,i}\in\{0,1\}$, the two vectors of covariates and the two disease status indicators for the two observations $i\in\{1,2\}$ of the pair. Then, the association between risk factors and subtype $k$ of the disease can be studied by applying a conditional logistic regression model, restricted to observations in stratum ${\cal M}_k$. Assume without loss of generality that data are arranged in such a way that the observation indexed $i=1$ is the case in each pair $\ell$, that is $Y^{(k)}_{\ell,1} = 1$ for all pairs $\ell$. Then, as usual under the conditional logistic regression model, we assume the existence of a vector $\bdelta_k^*\in\R^p$ such that the probability that the case is the one observed in pair $\ell$, given that a case is observed in pair $\ell$, writes \cite{greenland2000small}

\begin{equation}
 \P(Y^{(k)}_{\ell,1} = 1 | Y^{(k)}_{\ell,1} +Y^{(k)}_{\ell,2}  = 1, \bx^{(k)}_{\ell,1}, \bx^{(k)}_{\ell,2}) = \frac{1}{1+ \exp\{\bdelta_k^{*T} (\bx^{(k)}_{\ell,1} - \bx^{(k)}_{\ell,2})\}}\cdot \label{eq:modCondLog}
\end{equation}
 
 
Vector $\bdelta_k^*$ can then be estimated by maximizing the log conditional likelihood $\bdelta_k \longrightarrow L^{(cond)}_k(\bdelta_k)$ restricted to pairs in ${\cal M}_k$, which is defined as
\begin{align}
L^{(cond)}_k(\bdelta_k) 
&= - \sum_{\ell\in [m_k]} \log[1+ \exp\{\bdelta^T_k (\bx^{(k)}_{\ell,1} - \bx^{(k)}_{\ell,2}) \}] \nonumber\\
&= -[\log\{{\boldsymbol 1}_{m_k} + \exp(\bDelta^{(k)} \bdelta_{k} ) \}]^T{\boldsymbol 1}_{m_k} \label{eq:logVraisCond}
\end{align}
where ${\boldsymbol 1}_{m_k} = (1,\ldots, 1)^T \in \R^{m_k}$ and $\bDelta^{(k)}$ is the $m_k\times p$ matrix, whose $\ell$-th row corresponds to $(\bx^{(k)}_{\ell,1} - \bx^{(k)}_{\ell,2})$, for $\ell\in [m_k]$.

Equivalently, estimation of each $\bdelta^*_k$ can be performed simultaneously, though still independently, by maximizing the criterion
$
\sum_{k\in[K-1]} L^{(cond)}_k(\bdelta_k) =  -[\log\{{\boldsymbol 1}_{m} + \exp(\bar\bDelta  \bdelta ) \}]^T{\boldsymbol 1}_{m}
$
  over $\bdelta = (\bdelta^T_1, \ldots, \bdelta^T_{K-1})^T\in\R^{(K-1)p}$, with
$$
\bar\bDelta = \left( 
\begin{array}{c c c }
\bDelta^{(1)}& \hdots &{\bf 0}_{m_1,p}  \\
 \vdots & \ddots & \vdots \\
{\bf 0}_{m_{K-1},p}&\hdots  &  \bDelta^{(K-1)}  
\end{array}
\right).
$$ 

\subsection{Standard $L_1$ norm penalized estimation}\label{sec:CondLogit_Lasso}
Several packages have been developed to maximize a penalized version of criterion \eqref{eq:logVraisCond}: for instance, cLogitLasso is available within the R software \cite{Avalos};   the cLogitL1 package \cite{ReidTibsh} can also be used, although it is not maintained on the CRAN anymore. For appropriate values of the regularization parameter $\lambda$, they can be used to maximize  the following criterion over $\bdelta_{k} \in\R^p$,
$$ -[\log\{{\boldsymbol 1}_{m_k} + \exp(\bDelta^{(k)} \bdelta_{k} ) \}]^T{\boldsymbol 1}_{m_k}  - \lambda \|\bdelta_k\|_1$$
to get a sparse estimate of $\bdelta_k^*$. They can also be used to maximize the ``overall'' criterion
\begin{equation}  
-[\log\{{\boldsymbol 1}_{m} + \exp(\bar\bDelta  \bdelta ) \}]^T{\boldsymbol 1}_{m} - \lambda \|\bdelta\|_1 \label{LassoCond}
\end{equation}
over $\bdelta\in\R^{(K-1)p}$ to get a sparse estimate of $\bdelta^* = (\bdelta_1^*, \ldots, \bdelta_K^*)$. These two strategies are strictly identical and would return identical estimates. In particular, along both strategies, the estimation is performed independently on each stratum, that is independently for each subtype. This is likely sub-optimal when subtypes have commonalities. Indeed, these commonalities are expected to translate into some homogeneity among vectors  $\bdelta^*_{1}, \ldots, \bdelta^*_{K-1}$,  which may lead to improved estimation efficiency if properly accounted for.

\subsection{Data shared lasso}\label{sec:CondLogit_DSLasso}

Data shared lasso was independently proposed by Gross and Tibshirani (2016) \cite{Gross} and Ollier and Viallon (2017) \cite{AutoRefLasso} in the context of stratified regression models to account for the expected homogeneities among the parameter vectors to be estimated. The approach relies on the following over-parametrized decomposition for each parameter $\delta^*_{k,j}$, for $k\in[K-1]$ and $j\in[p]$, 
\begin{equation}
 \delta^*_{k,j} = \mu^*_j + \gamma^*_{k,j}. \label{eq:decompDSlasso}
 \end{equation}
Here $\mu^*_j$ can be seen as the ``global'' parameter for covariate $j$ and is common to all subtypes, while $\gamma^*_{k,j}$ captures the variation of the parameter for subtype $k$ around this global parameter.  Even if decomposition \eqref{eq:decompDSlasso} is over-parametrized, estimates of  $\mu^*_j$ and $\gamma^*_{k,j}$ for $k\in[K-1]$ and $j\in[p]$ can be derived by maximizing the following criterion over $\bmu = (\mu_1, \ldots, \mu_p)$ and the $\bgamma_k$'s, with $\bgamma_k= (\gamma_{k,1}, \ldots, \gamma_{k,p})$,
\begin{align}
& \sum_{k\in[K-1]} L^{(cond)}_k(\bmu + \bgamma_k)  - \lambda (\|\bmu\|_1 + \sum_{k=1}^{K-1} \tau_k \|\bgamma_k\|_1). \label{eq:DScrit}
\end{align}
The $L_1$-norm penalty $\|\bmu\|_1$ encourages sparsity of the vector of global parameters, while $\|\bgamma_k\|_1$ encourages homogeneity among vectors  
$\hbdelta_k$ defined as $\hbdelta_k = \hbmu + \hbgamma_k$, for $k\in[K-1]$. For appropriate values of the regularization parameters $\lambda$ and $\tau_k$, data shared lasso allows the estimation of parameters $\hmu_j$ under one of the infinitely many  decompositions of the form  (\ref{eq:decompDSlasso}). 
Any particular choice for $\tau_k$ leads to a particular ``definition'' of the estimated global parameter $\hmu_j$ for covariate $j$. Given this particular definition, data shared lasso returns estimates $\hbdelta_1, \ldots, \hbdelta_{K-1}$ that are typically close to $\hbmu=(\hmu_1, \ldots, \hmu_p)$ in the $L_1$-norm sense. For instance, if $\tau_k=1$ for all $k\in[K-1]$, it showed that\cite{Gross,AutoRefLasso}
$$ \hmu_j  = \underset{m}{{\rm argmin}} \{|m| + \sum_{k\in[K-1]}|\hdelta_{k,j}-m|\} = {\rm median}(\hdelta_{1,j}, \ldots, \hdelta_{K-1,j}, 0).$$

In other respect, several more standard approaches turn out to be special cases of data shared lasso. If $\tau_k=\infty$ for all $k$, then $\hbdelta_k = \hbmu$ for all $k\in[K-1]$ and data shared lasso reduces to the approach that consists in pooling all strata together; we will refer to this strategy as ``Pooled''. ``Pooled'' overlooks the subtype specificities and generally leads to biased estimates of vectors $\bdelta^*_k$. On the other hand, for large enough values of $\lambda$, we have $\hbmu = {\bf 0}$ and for appropriate values of parameters $\tau_k$, data shared lasso reduces to estimating each vector $\bdelta^*_k$ independently just as in \eqref{LassoCond} above; we will refer to this strategy as ``Indep''. ``Indep'' overlooks the commonalities among the subtypes, hence typically leads to estimates with unnecessarily high variance. Finally, setting $\tau_r = \infty$ for one particular $r\in[K-1]$ corresponds to working under the constraint $\hbbeta_r = \hbmu$. In this case, data shared lasso reduces to another standard approach which consists in first selecting subtype $r$ as a reference, and then including interaction terms between each covariate and the indicator variables $\1(S=k)$ for $k\neq r$; we will refer to this strategy as ``Ref''. Note that, for any particular choice $r$, the model complexity $C(r)$ is naturally defined as the number of non-zero parameters to be estimated, that is $C(r) = \|\bbeta^*_r\|_0 + \sum_{k\neq r} \|\bbeta^*_k-\bbeta^*_r\|_0$, with $\|\cdot\|_0 $ standing for the $L_0$ pseudo-norm. Consequently, the model complexity and estimation efficiency of ``Ref'' critically depend on the arbitrary choice of the reference stratum, that is the reference subtype in our case; see \cite{AutoRefLasso} for more details. Data shared lasso by-passes this arbitrary choice and, under stratified linear regression models, was shown to perform nearly as well as the oracular (and inapplicable) version of ``Ref''  based on an optimal and covariate-specific choice for the reference stratum. 

Another nice property of data shared lasso is that it is readily implementable given any standard lasso solver. In particular, the data shared lasso criterion \eqref{eq:DScrit} above can be rewritten as
\begin{align*}
& \sum_{k\in[K-1]} L^{(cond)}_k(\bmu + \bgamma_k)  - \lambda (\|\bmu\|_1 + \sum_{k=1}^{K-1} \tau_k \|\bgamma_k\|_1) \\
&= -[\log\{{\boldsymbol 1}_{m} + \exp(\tilde\bDelta_{\btau}  \bGamma ) \}]^T{\boldsymbol 1}_{m} - \lambda \|\bGamma\|_1 
\end{align*}
with $\btau=(\tau_1, \ldots, \tau_{K-1})$, $\bGamma=(\bmu^T, \tilde\bgamma_1^T, \ldots, \tilde\bgamma^T_{K-1})$ and 
$$
\tilde\bDelta_{\btau} = \left( 
\begin{array}{c c c c }
\bDelta^{(1)}&\frac{\bDelta^{(1)}}{\tau_1}& \hdots &{\bf 0}_{m_1,p}  \\
\vdots & \vdots & \ddots & \vdots \\
 \bDelta^{(K-1)} &{\bf 0}_{m_{K-1},p}&\hdots  &  \frac{\bDelta^{(K-1)}}{\tau_{K-1}}  
\end{array}
\right).
$$ 
This criterion is exactly of the same form as \eqref{LassoCond}: as a result, running cLogitLasso with the design matrix $\tilde\bDelta_{\btau}$ returns a vector  $(\hbmu^T, \hat{\tilde \bgamma}_1^T, \ldots,\hat{\tilde \bgamma}^T_{K-1})$ from which a data shared lasso estimates $\hbdelta_k = \hbmu + \hat{\tilde \bgamma}_k /\tau_k$ can be derived for $\bdelta^*_k$, $k\in[K-1]$. 

We will illustrate the performance of data shared lasso when analyzing matched case-control studies with multiple subtypes of cases through simulated examples in Section \ref{sec:Simul}, as well as through the analysis of a case-control study for breast cancer nested in the EPIC cohort in Section \ref{sec:App}.

\section{Unmatched case-control studies with multiple subtypes of cases and sparse multinomial logistic models}
\label{sec:UnMatched}

We now turn our attention to the unmatched setting. When $K-1$ subtypes of cases are present for some given integer $K>1$, the outcome $Y$ can be modeled as a categorical variable, taking values in $[K]$. Hereafter, we will assume that $Y=K$ for controls, while $Y=k$ for cases of subtype $k$, for any $k\in[K-1]$. When no natural order exists among the categories of $Y$, the multinomial logistic regression model is a natural extension of the standard logistic regression model. Below, we will recall some basics about the multinomial logistic regression model. In particular, we will present two formulations of this model under which $L_1$-norm penalized estimation can be performed.  We will then establish the relationship  between these two approaches, basing our arguments on the data shared lasso ideas. 

\subsection{The multinomial logistic regression model}
For ease of notation, we will mostly focus on models with no intercept. Then, in its symmetric formulation, the multinomial logistic regression model assumes the existence of $K$ vectors $(\bbeta^*_1, \ldots, \bbeta^*_K)\in(\R^p)^K$ such that 
\begin{align}
\P(Y=k| \bx=\bx_0) =\frac{\exp(\bx_0^T \bbeta_k^*)}{\sum_{\ell=1}^K \exp(\bx_0^T \bbeta_\ell^*)},\label{eq:ModSym}
\end{align}
for any value $\bx_0\in\R^p$ of the covariate vector, with $p\geq 1$. Because $\sum_{k\in[K]} \P(Y=k | \bx=\bx_0)  = 1$ for any $\bx_0\in\R^p$, this formulation is over-parametrized and vectors $\bbeta^*_1, \ldots, \bbeta^*_K$ in Equation (\ref{eq:ModSym}) are defined up to a constant only. More precisely, if model (\ref{eq:ModSym}) holds with vectors $\bbeta^*_1, \ldots, \bbeta^*_K$, then it holds with vectors $\bbeta^*_1 +\bnu, \ldots, \bbeta^*_K+\bnu$ as well, for any $ \bnu\in \R^p$. 

To resolve this identifiability issue, a standard solution consists in selecting a reference category, say the $K$-th one without loss of generality. This leads to the constraint $\bbeta^*_K = {\bf 0}_p$ in the formulation above, and the multinomial logistic regression model  then reduces to assuming the existence of  $(\bdelta^*_1, \ldots, \bdelta^*_{K-1})\in(\R^{p})^{K-1}$ such that 
\begin{align}
\P(Y=k | \bx=\bx_0) =\frac{{\exp(\bx_0^T \bdeltak^*)}^{\1(k\neq K)}}{1 + \sum_{\ell=1}^{K-1} \exp(\bx_0^T \bdelta_\ell^*)} 
. \label{eq:ModRefK}
\end{align}
Of course, the two formulations are strictly equivalent and from any ``initial'' vectors of parameters $\bbeta^*_1, \ldots, \bbeta^*_K$ satisfying Equation (\ref{eq:ModSym}),  Equation (\ref{eq:ModRefK}) holds with vectors $\bdelta^*_1, \ldots, \bdelta^*_{K-1}$ defined as $\bdelta^*_k = \bbeta^*_k - \bbeta^*_K$, for $k\in[K]$. 

Vectors $\bdelta^*_1, \ldots, \bdelta^*_{K-1}$ in Equation \eqref{eq:ModRefK} can be estimated through likelihood maximization. Assume the data consists of $n$ independent and identically distributed replica $(\bx_i, Y_i)_{1\leq i\leq n}$ with $\bx_i\in\R^p$ and $Y_i\in[K]$. Then, under model (\ref{eq:ModRefK}), the log-likelihood is defined for any $(\bdelta_1, \ldots, \bdelta_{K-1})\in(\R^p)^{K-1}$ as
\begin{align}
L(\bdelta_1, \ldots, \bdelta_{K-1}) &= \frac{1}{n}\sum_{i=1}^n \log\big( p_{y_i}(\bx_i; \bdelta_1, \ldots, \bdelta_{K-1}, {\bf 0}_p)\big)\label{eq:logVraisPolyt}
\end{align}
where, for any collection of vectors $(\bu_1, \ldots, \bu_K)\in\R^{p\times K}$, we set 
\begin{align}
p_\ell(\bx_0; \bu_1, \ldots, \bu_K) = \frac{\exp(\bx_0^T \bu_\ell)}{\sum_{k=1}^K \exp( \bx_0^T \bu_k)}\cdot
\end{align}

\subsection{Sparse estimation under the standard formulation}
A first sparse approach, that will be referred to as MultinomSparseRef here, simply consists in maximizing the $L_1$-norm penalized version of the log-likelihood defined in \eqref{eq:logVraisPolyt} 
\begin{align}
\max_{\bdelta_1, \ldots, \bdelta_{K-1}} \left[  L(\bdelta_1, \ldots, \bdelta_{K-1}) - \lambda \sum_{k=1}^{K-1} \|\bdelta_k\|_1 \right].
\label{eq:ModRefK_L1pen}
\end{align}
Maximizers of (\ref{eq:ModRefK_L1pen}) can be obtained via the algorithm described in \cite{Krishnapuram}. Thanks to the well-known link between the log-likelihood $L$ and the conditional logistic log-likelihood \cite{hendrickx2000special}, they can also be obtained as solutions returned by package cLogitLasso \cite{Avalos} after a simple modification of the original data.

\subsection{Sparse estimation under the symmetric formulation}

Package glmnet in R \cite{glmnet} implements an $L_1$-penalized approach based on the symmetric formulation of the model, which will be referred to as MultinomSparseSym here. Parameter vectors $\bbeta^*_1, \ldots, \bbeta^*_K$ used under formulation (\ref{eq:ModSym}) cannot be estimated by standard maximum likelihood estimation because of the aforementioned  lack of identifiability. But because penalizing acts as constraining, estimates of $\bbeta^*_1, \ldots, \bbeta^*_K$ can be obtained as maximizers of the $L_1$-penalized version of the  following log-likelihood $$\cL(\bbeta_1, \ldots, \bbeta_K) = \frac{1}{n}\sum_{i=1}^n \log\{ p_{y_i}(\bx_i; \bbeta_1, \ldots, \bbeta_K)\}.$$ More precisely, package glmnet maximizes the following criterion over $(\bbeta_1, \ldots, \bbeta_K)\in{\R^p}^K$,
\begin{align*}
  \cL(\bbeta_1, \ldots, \bbeta_K) - \lambda \sum_{k=1}^K \|\bbeta_k\|_1
\end{align*}
 for some appropriate value of the regularization parameter $\lambda$. In \cite{glmnet}, it is shown that maximizers $\hbbeta_1, \ldots, \hbbeta_K$ of this criterion are such that 
\begin{equation}
{\rm median}(\hbeta_{1, j}, \ldots, \hbeta_{K,j}) = 0, \quad{\rm for\ all\ }j\in[p]. \label{eq:medianL1sym}
\end{equation}
See the Appendix for an alternative proof of this result. Equation \eqref{eq:medianL1sym} establishes that penalizing by the $L_1$-norm under the symmetric formulation of the model implicitly solves the lack of identifiability for each covariate by constraining the median of its parameters across the $K$ categories to be null.  

We shall recall that when intercepts are considered, as is often the case in practice, they are generally not penalized. Setting $\bbeta_k^T = (\beta_{k,0}, \bbeta_{k,\setminus 0}^T)$ where $\beta_{k,0}$ stands for the intercept term for the $k$-th category, the penalty term then becomes $\lambda \sum_{k=1}^K \|\bbeta_{k,\setminus 0}\|_1$. Then, identifiability issues are still present for the intercept terms under the symmetric formulation of the model. In the glmnet package, this is resolved by mean centering, which corresponds to imposing the constraint $\sum_{k\in[K] } \hat \beta_{k,0} = 0$ \cite{glmnet}.

\subsection{Relationship between MultinomSparseSym and MultinomSparseRef}\label{sec:CompSparseSymRef}

Consider the standard formulation \eqref{eq:ModRefK}. When $Y=K$ stands for controls and $Y=k$ stands for cases of subtype $k$ for $k\in[K-1]$ , some level homogeneity among vectors  $\bdelta^*_{1}, \ldots, \bdelta^*_{K-1}$ is often expected. The ideas of data shared lasso can then be applied. When combined with MultinomSparseRef, data shared lasso first consists in considering the decomposition $\delta^*_{k,j} = \mu^*_j + \gamma^*_{k,j}$
for $k\in[K-1]$ and $j\in[p]$, and then maximizing the following criterion, over $\bmu = (\mu_1, \ldots, \mu_p)$ and the $\bgamma_k$'s, with $\bgamma_k= (\gamma_{k,1}, \ldots, \gamma_{k,p})$,
\begin{align*}
L(\bmu + \bgamma_1, \ldots,\bmu + \bgamma_{K-1})  - \lambda (\|\bmu\|_1 + \sum_{k=1}^{K-1} \|\bgamma_k\|_1).
\end{align*}
We will refer to this approach as MultinomDataSharedRef hereafter. This criterion is of the same form as \eqref{eq:DScrit}, for the particular choice $\tau_k=1$ for all $k\in[K-1]$. Now, denote by $(\hbbeta_1, \ldots, \hbbeta_K)$ and $(\hbmu, \hbgamma_1, \ldots,\hbgamma_{K-1})$ the solutions returned by  MultinomSparseSym and MultinomDataSharedRef, respectively. In the Appendix we show that $\hbmu = - \hbbeta_K$ and $\hbbeta_k = \hbgamma_k$ for all $k\in[K-1]$. This result formally establishes the equivalence between MultinomDataSharedRef and MultinomSparseSym: working under formulation (\ref{eq:ModSym}) with an $L_1$-norm penalty, as implemented in the glmnet package, exactly corresponds to working under formulation (\ref{eq:ModRefK}) with a data shared lasso penalty (for the particular choice $\tau_k = 1$ for all $k\in[K-1]$) to encourage homogeneity among vectors $(\bdelta^*_1, \ldots, \bdelta^*_{K-1})$. 

To get a better understanding of the relationship between MultinomSparseSym and MultinomSparseRef, denote by $(\tilde \bbeta_1, \ldots, \tilde \bbeta_K)$ maximizers of the criterion
$\LL(\bbeta_1, \ldots, \bbeta_K) - \lambda ( \|\bbeta_K\|_1 + \sum_{k\in[K-1]}  \|\bbeta_k - \bbeta_K\|_1)$. In the Appendix, we show that $\tilde \bbeta_K = {\bf 0}_p$ and $\tilde \bbeta_k = \hbdelta_k$ for $k\in[K-1]$, where $\hbdelta_k$ are estimates returned by MultinomSparseRef, that is maximizers of 
$L(\bdelta_1, \ldots,\bdelta_{K-1})  - \lambda  \sum_{k=1}^{K-1} \|\bdelta_k\|_1$. Therefore, applying MultinomSparseRef after selecting the $K$-th category as the reference corresponds to working under the symmetric formulation and encouraging similarities between $\hbbeta_K$ and the other vectors $\hbbeta_k$ for $k\in[K-1]$. This strategy is expected to perform best if ${\cal C}(K) =  \sum_{k\in[K]}\|\bbeta^*_k - \bbeta^*_K\|_0$ is small. In other words, while the choice of the reference category has no effect whatsoever when estimation is done by maximizing the unpenalized log-likelihood \eqref{eq:logVraisPolyt}, this choice is critical for MultinomSparseRef, that is when the  $L_1$-penalized log-likelihood \eqref{eq:ModRefK_L1pen} is maximized. This is closely related to our discussion about the performance of the ``Ref'' strategy described in Section \ref{sec:CondLogit_DSLasso} under matched designs (and more generally under stratified regression models), which also critically depends on the arbitrary choice for the reference stratum. For illustration, consider the following toy example where $(i)$ $\bbeta^*_1=\cdots =\bbeta^*_{K-1}$, and $(ii)$ $\beta^*_{K,j}\neq \beta^*_{1,j}$ for all $j\in[p]$. When $Y=K$ indicates controls while $Y=k$ for $k\in[K-1]$ indicates cases of subtype $k$, this situation arises when all subtypes are actually identical. Then we have ${\cal C}(K) = (K-1)p$ while ${\cal C}(r) = p$ for any $r<K$, with ${\cal C}(r) = \sum_{k\in[K]} \|\bbeta^*_k - \bbeta^*_r\|_0$ standing for the model complexity when setting the reference category to $r$ before applying MultinomSparseRef. In this example, category $K$ is the worst choice for the reference when using MultinomSparseRef, even if it would be regarded as the most natural choice by many practitioners. 

\section{Simulation study}\label{sec:Simul}
\subsection{The matched setting}\label{sim:SimulMatched}
We performed a simulation study to assess the performance of data shared lasso in the context of matched case-control studies when $(K-1)=6$ subtypes of cases are present. We compared it with two more standard strategies: Indep and Ref. For the latter, the first subtype was selected as the subtype of reference. In addition, we implemented a cross-validation technique similar in spirit to the one-step lasso \cite{buhlmann2008discussion}  to select optimal regularization parameters and obtain final parameter estimates. To save computational time, data shared lasso and Ref were implemented with one particular choice for $\tau_k$ only, that is $\tau_k = 1$ for all $k\in[K-1]$. 

We set the number of observations to $n=1000$ and the number of covariates was set to $p=100$. Covariates were randomly generated under a multivariate Gaussian distribution ${\cal N}({\bf 0}_p, \bSigma)$, where $\bSigma_{i,j} = (0.25^2)\times0.3^{|i-j|}$. Pairs of observations were then created and randomly assigned to one stratum ${\cal M}_k$ in such a way that $m_1=200$, $m_2=100$ and $m_k=50$ for $k=3, \ldots, 6$. Within each pair $\ell$ of each stratum ${\cal M}_k$, the response variable $Y^{(k)}_{\ell, 1}$ was then generated according to Equation (\ref{eq:modCondLog}), while $Y^{(k)}_{\ell, 2}$ was set to $1-Y^{(k)}_{\ell, 1}$. As for parameters $\delta^*_{k,j}$, they were defined as follows. One subset $J_1\subset[p]$ was first randomly selected, with $|J_1|=10$. For $j\notin J_1$, we set $\delta^*_{k,j}=0$ for all $k\in[K-1]$. For $j\in  J_1$, four configurations were considered, allowing the level of homogeneity among $(\delta^*_{1,j}, \ldots, \delta^*_{K-1,j})$ to vary. In the first configuration (full homogeneity), we set $\delta^*_{k,j}= \iota_j \delta$, for some $\delta >0$ and with $\iota_j = \pm 1$. In the second configuration (weak heterogeneity), for $j\in J_1$, we randomly select one $k_j\in [K-1]$, set $\delta^*_{k,j} =\iota_{k,j} \delta$ for $k\neq k_j$ and $\delta^*_{k_j, j}=\iota_{k_j,j} \delta*(1+U_{k_j,j})$, with each $\iota_{k,j}=\pm1$ and $U_{k_j,j}\sim {\cal U}_{[\sqrt{K}/2, 2\sqrt{K}]}$.  In the third configuration (moderate heterogeneity), we randomly select three indices $(k_{j,1}, k_{j,2}, k_{j,3})\in [K-1]^3$, set $\delta^*_{k,j} =\iota_j \delta$  $k\notin \{k_{j,1}, k_{j,2}, k_{j,3}\}$ and  $\delta^*_{k, j}=\iota_{k,j} \delta*(1+U_{k,j})$ for $k \in \{k_{j,1}, k_{j,2}, k_{j,3}\}$, with again $\iota_{k,j}=\pm1$ and $U_{k,j}\sim {\cal U}_{[\sqrt{K}/2, 2\sqrt{K}]}$. Finally, in the fourth configuration (full heterogeneity), we set $\delta^*_{k, j}=\iota_{k,j} \delta*(1+U_{k,j})$ for $k\in[K-1]$ with again $\iota_{k,j}=\pm1$ and $U_{k,j}\sim {\cal U}_{[\sqrt{K}/2, 2\sqrt{K}]}$. In each configuration, parameter $\delta$ varied in $\{0.4, 1.0, 2.0, 3.0\}$ to study the impact of signal strength on the performance of the approaches.  

One simulation design here corresponds to one particular combination of the value for $\delta$ and the level of heterogeneity. Fifty replications of each simulation design were performed and results presented below correspond to averages of the considered criteria over these 50 replicates for each approach.  

\begin{figure}
\includegraphics[scale=0.7]{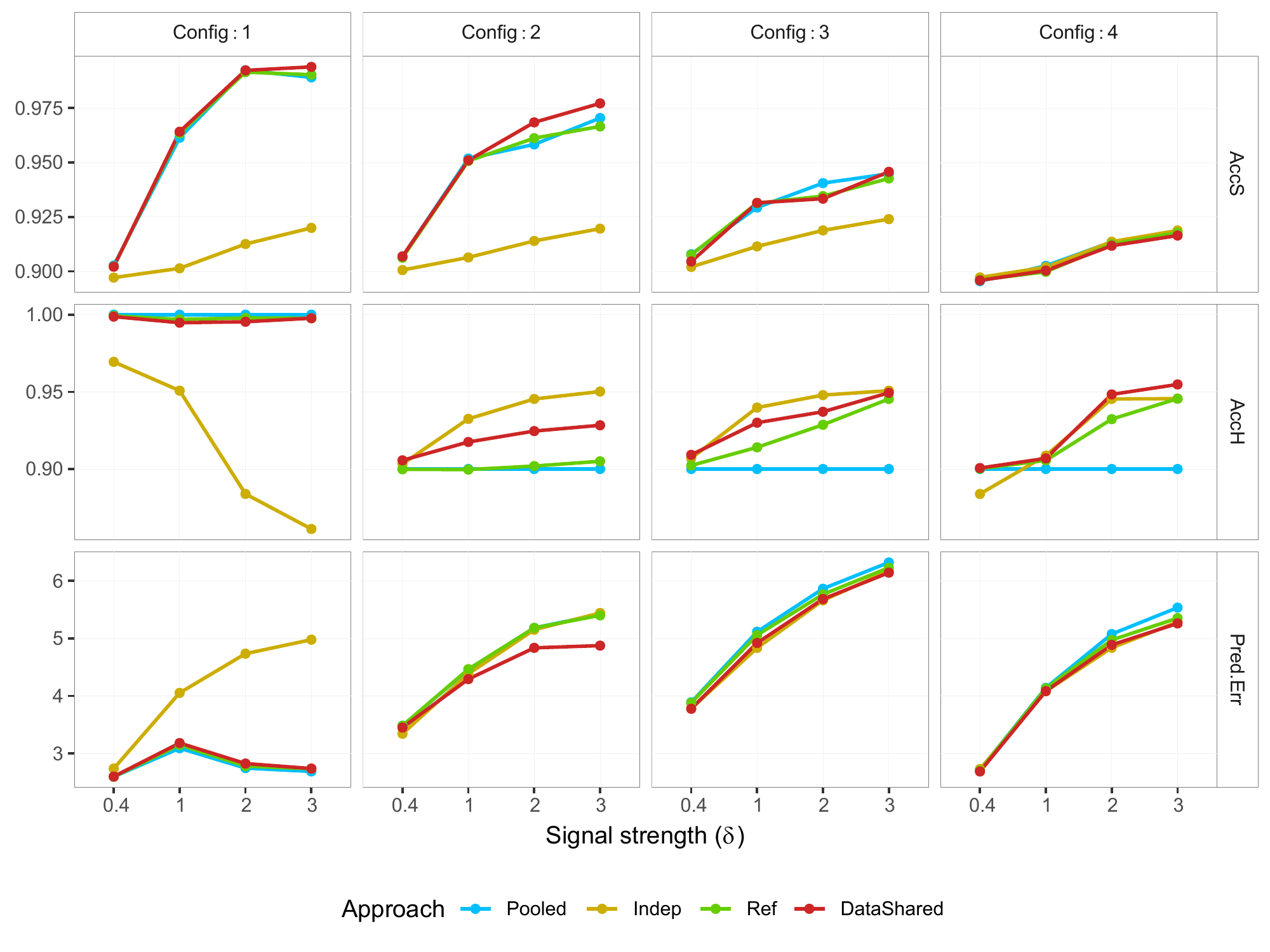}
\caption{Results of the simulation study in the matched setting.}\label{fig:Simul_Matched}
\end{figure}

Figure \ref{fig:Simul_Matched} presents the results regarding support recovery of the parameter matrix ${\bf D}^* = (\bdelta^*_1, \ldots, \bdelta^*_{K-1})\in\R^{p\times (K-1)}$  (AccS; the higher, the better), the identification of heterogeneities among vectors $(\delta^*_{1,j}, \ldots, \delta^*_{K-1,j})$, for $j\in[p]$  (AccH; the higher, the better), as well as prediction error (Pred.Err; the lower, the better). 
Overall, the performance of Indep does not depend on the level of homogeneity, while those of DataShared and Ref typically increase with the homogeneity level. This was expected since Indep does not account for homogeneity, while DataShared and Ref do. In case of full homogeneity among vectors $\bdelta_1^*, \ldots,\bdelta_{K-1}^*$ (Configuration 1), Ref and DataShared perform similarly regarding the three criteria, they perform as well as Pooled, and clearly outperform Indep. The similar performance of Ref and DataShared was expected in this particular case where model complexity $C(r)$ defined in Section \ref{sec:CondLogit_DSLasso} does not depend on $r$. In case of full heterogeneity (Configuration 4), DataShared and Ref again perform similarly, as expected since  $C(r)$ still does not depend on $r$. Of course, they do not perform better than Indep in this case, but it is noteworthy that they do not perform worse either.  In configurations 2 and 3 (weak and moderate heterogeneities), data shared lasso generally leads to the best results regarding prediction error and, to a lesser extent, support recovery and identification of the heterogeneities. In particular, it outperforms Ref, which confirms that by by-passing the arbitrary choice of the reference category, data shared lasso generally better accounts for homogeneity than Ref does when such homogeneity exists.  These results are consistent with those obtained when evaluating data shared lasso under linear regression models \cite{AutoRefLasso}  and binary graphical models \cite{NadimBinGraph}. 

\subsection{The unmatched setting}\label{sim:SimulUnMatched}

We also performed a simulation study in the unmatched setting to illustrate the relative interests of MultinomSparseRef and MultinomSparseSym (the later being the same as MultinomDataSharedRef) depending on the level of homogeneity among vectors $\bdelta^*_{1}, \ldots, \bdelta^*_{K-1}$ under formulation \eqref{eq:ModRefK} or, equivalently, among vectors $\bbeta^*_{1}, \ldots, \bbeta^*_{K}$ under formulation \eqref{eq:ModSym}. Again, we chose $K=7$. To save computational times, and because conclusions were consistant with those drawn in the matched case, a low-dimensional setting with $n=1000$ and $p=20$ was considered here. For data generation, we adapted the framework described in Section \ref{sim:SimulMatched} to the unmatched setting using formulation \eqref{eq:ModRefK}. We used intercept terms, $(\delta_{0,1}, \ldots, \delta_{0, K-1})$, chosen in such a way that $\P(Y=K)=0.5$ and $\P(Y=k)$ ranged from 0.05 to 0.2 for $k\in[K-1]$. In this low-dimensional setting, regularization parameters were selected as minimizers of the BIC after adapting the Lasso-OLS hybrid ideas to our context \cite{lars}.

\begin{figure}
\includegraphics[scale=0.7]{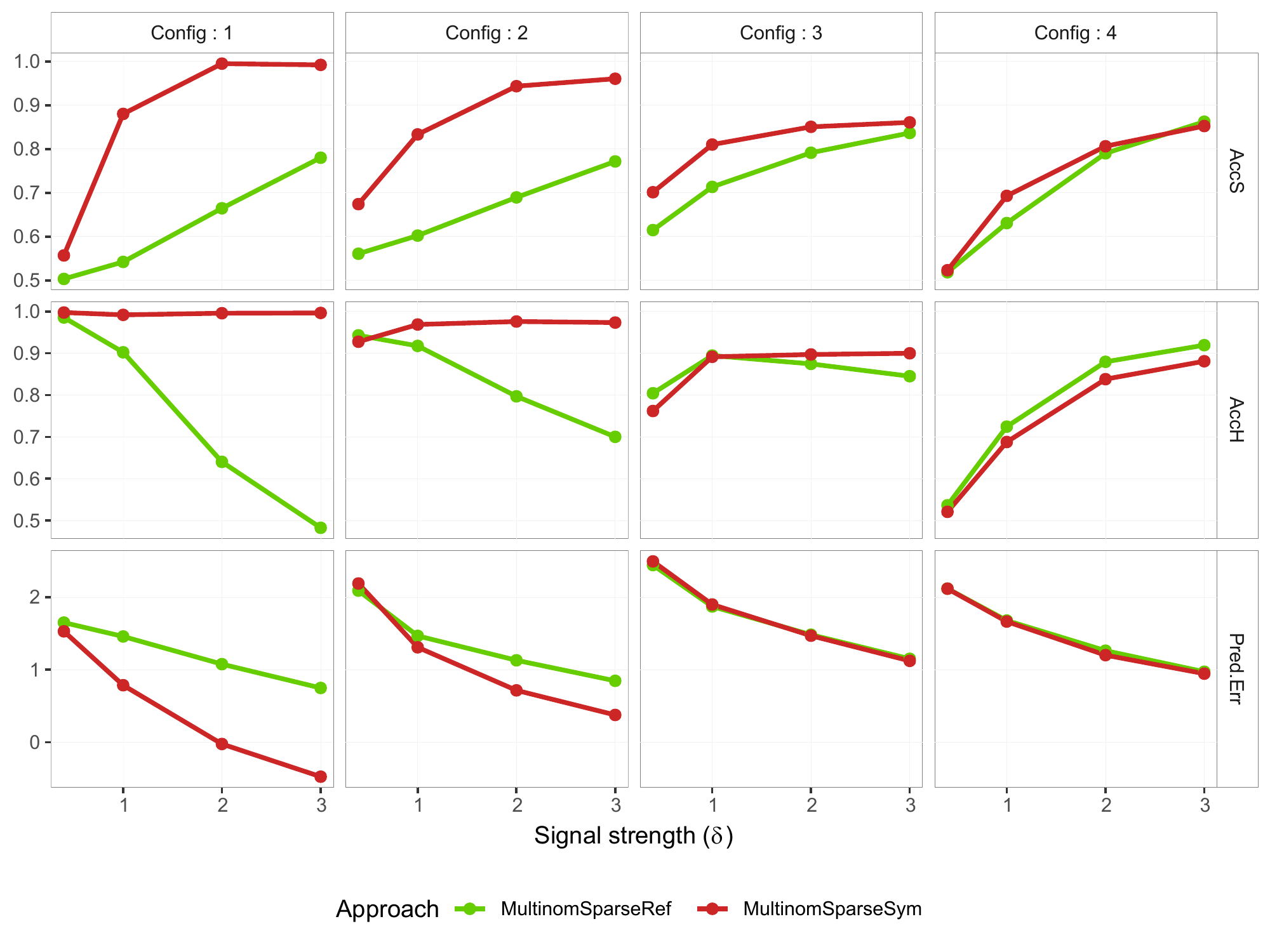}
\caption{Results of the simulation study in the unmatched setting.}\label{fig:Simul_Unmatched}
\end{figure}

Figure \ref{fig:Simul_Unmatched} presents the results in this unmatched setting. They confirm that using data shared lasso (or, equivalently, the symmetric formulation) allows the homogeneity to be accounted for when present, which translates into better predictive performance, support recovery and identification of the heterogeneties. We shall also stress that even in the case of full heterogeneity, MultinomSparseSym performs as well as MultinomSparseRef, just as data shared lasso and Ref did in the matched setting case.

\begin{figure}
\centering
\includegraphics[scale=0.45]{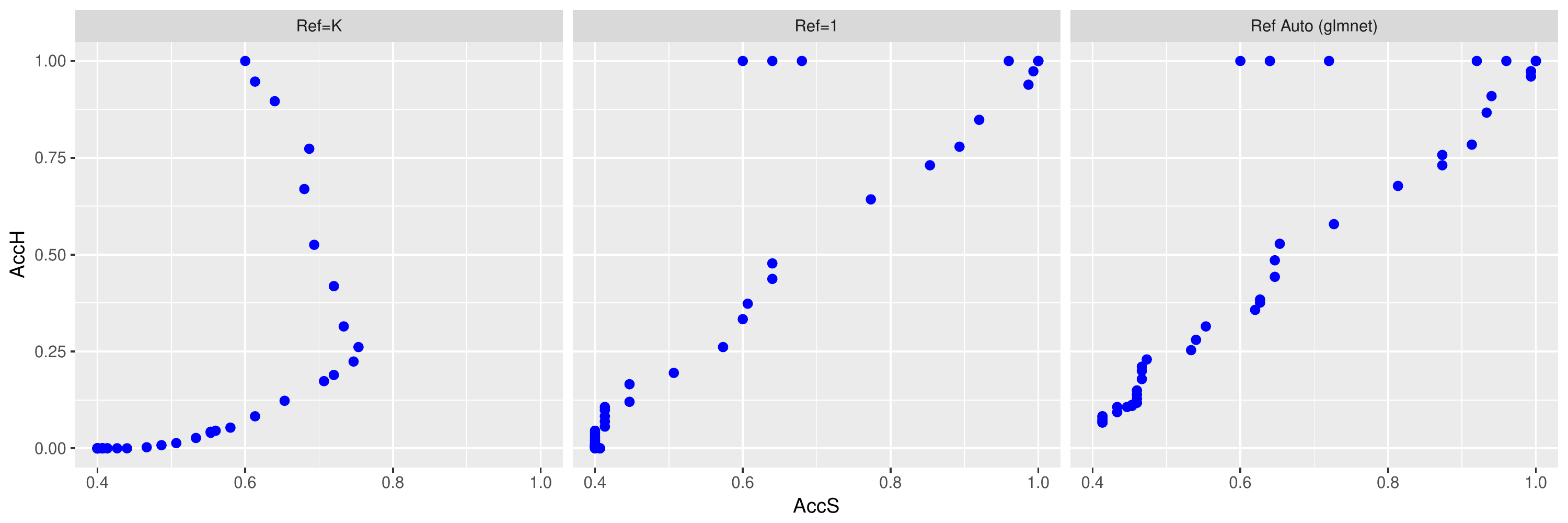}
\caption{Comparison of the results obtained with MultinomSparseRef for two distinct choices of the reference category and MultinomSparseSym, in the case of full homogeneity among vectors $(\bdelta^*_1, \ldots, \bdelta^*_{K-1})$.}\label{fig:Comp_SparseRef1_RefK}
\end{figure}

We further investigated in more details the poor performance of MultinomSparseRef. We focused on the particular case of full homogeneity among vectors $\bdelta^*_{1}, \ldots, \bdelta^*_{K-1}$ under formulation \eqref{eq:ModRefK}.  For one sample generated under configuration one (full homogeneity) with $\delta=3$ (corresponding to a large signal strength), we computed criteria AccS and AccH for the sequence of parameter vectors estimates returned by MultinomSparseRef and MultinomSparseSym for varying values of the regularization parameter $\lambda$ on appropriate grids $[\lambda_1/1000, \lambda_1]$. Here $\lambda_1$ was set, as usual, as the minimal value for which the considered method returned a null parameter vector. MultinomSparseRef was actually ran with two particular choices for the reference category. We primarily chose category $K$ as in Figure \ref{fig:Simul_Unmatched}. We recall that this choice is quite natural when category $K$ corresponds to controls. We also recall that in this case of full homogeneity among vectors $\bdelta^*_{1}, \ldots, \bdelta^*_{K-1}$,  we have ${\cal C(K)} = 10(K-1)=60$ while ${\cal C}(r)=10$ for any $r\neq K$. We then also implemented  MultinomSparseRef  with reference category set to $1$. Results returned by these two versions of MultinomSparseRef were compared to those returned by MultinomSparseSym (or equivalently, MultinomDataSharedRef). In each panel of Figure \ref{fig:Comp_SparseRef1_RefK}, each point represents values for AccS ($x$-axis) and AccH ($y$-axis) over the grid of regularization parameters used for the corresponding method. The choice of controls as the reference category (left panel, Ref$=K$), though standard, prevents MultinomSparseRef from visiting models with AccS greater than 0.75 whatever the value of the regularization parameter. On the other hand, choosing any subtype of cases as the reference (center panel, Ref$=1$ here) allows MultinomSparseRef to visit models with higher values for both AccS and AccH. Models visited by MultinomSparseSym are very similar to those visited by MultinomSparseRef with the optimal reference category. These results confirm that $(i)$ the performance of MultinomSparseRef critically depends on the arbitrary choice of the reference category when homogeneity is high, and $(ii)$ MultinomSparseSym (resp., equivalently, MultinomSparseRef with a data shared lasso penalty) by-passes (resp. corrects)  the arbitrary choice of the reference category,  and allows the visit of nearly the same models as those visited when applying MultinomSparseRef with the optimal choice for the reference category.

\section{Application}\label{sec:App}
\subsection{The data}
The European Prospective Investigation into Cancer and Nutrition (EPIC) study is an ongoing multicenter prospective study aiming to investigate prospectively the etiology of cancer in relation to diet, lifestyle and environmental factors, and for which the study design have been previously describe in detail \cite{riboli2002european}.  From 1992 to 2000, a total of 521,324 participants were recruited across 10 European countries, mostly from the general population, of which 70\% are women, aged from 35 to 70 years. Among these participants, 246,000 women provided a blood sample at inclusion.
Here, we present  preliminary results from the analysis of a case-control study nested in EPIC, whose main objective was to assess the association between metabolites and the risk of subtypes of breast cancer. 1635 cases of breast cancer were included, along with 1635 matched controls (using incidence density sampling). For all these individuals,  plasma samples collected at inclusion in the study were analyzed by mass spectrometry (AbsoluteIDQ p180 Kit) allowing the measurement of the levels of 127 metabolites. Those metabolites have been anonymized here since biological interpretation is out of the scope of this preliminary analysis. We considered six subtypes for cases, based on the presence/absence of hormone receptors: HER2-enriched, triple negative, Luminal A PR+, Luminal A PR-, Luminal B PR+ and Luminal B PR-.  

\subsection{Results}
\begin{figure}
\includegraphics[scale=0.8]{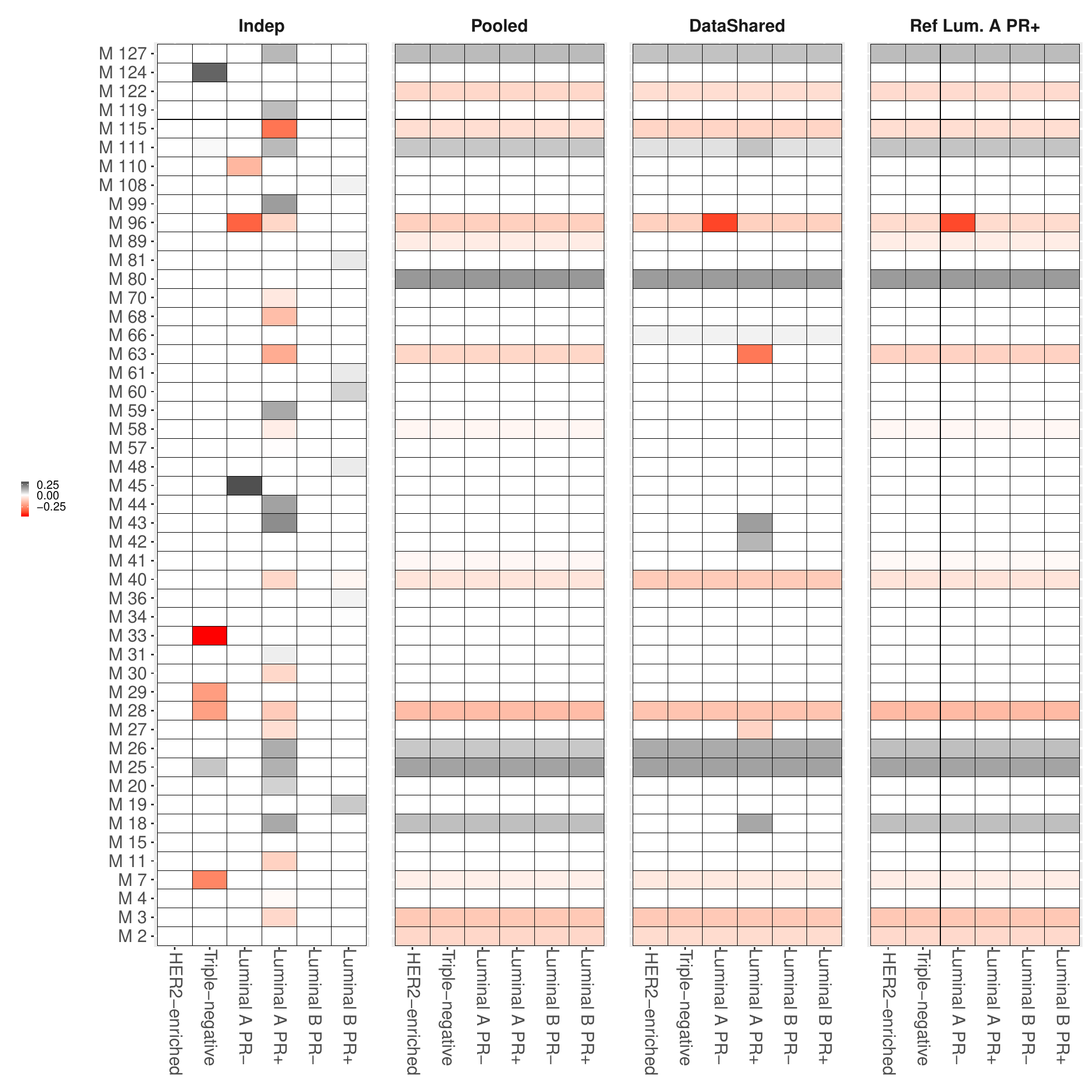}
\caption{Preliminary results from the analysis of the matched case-control study nested in EPIC.}\label{fig:Appli}
\end{figure}

We estimated sparse conditional logistic regression models based on the Indep, Pooled, Ref and data shared lasso strategies described in Section \ref{sec:Matched}. For the Ref strategy, Luminal A PR + was chosen as the reference subtype, which we believe would be considered as a natural choice by most practitioners because it is the most common subtype. Results are presented in Figure \ref{fig:Appli}, where only metabolites identified as potential predictor of at least one breast cancer subtype by at least one approach have been retained. As expected, using either Ref or data shared lasso lead to much more interpretable results than the Indep and Pooled strategies when the objective is to identify potential heterogeneities across subtypes. Data shared lasso allows the identification of a few heterogeneities, in particular for the the most common subtype, Luminal A PR+. Interestingly, Ref was not able to identify any heterogeneities for this subtype: this is because it was used as the reference subtype. We shall however mention that no notable difference was observed in terms of prediction errors when comparing the models returned by Pooled, Ref and Data Shared Lasso (Indep was slightly worse than its competitors). This can be explained by the fact that the association between the metabolites 
and subtypes of breast cancer is rather limited. We still believe that this application nicely illustrates the potential benefit of the data shared lasso strategy which may help hierarchize the most probable heterogeneities between subtypes: in the present example, M96 might be of particular interest for Luminal A PR-, while M18, M27, M42, M43, M63 and M111 might be specific to Luminal A PR+.

\section{Discussion}\label{sec:Discussion}

We considered the analysis of case-control studies when several subtypes of cases exist, which is increasingly common in cancer epidemiology. Considering both matched and unmatched settings, we showed that data shared lasso was a simple approach, which accounts for commonalities among the subtypes, when present, and improves estimation efficiency. In the unmatched setting, our observations provide practical guidance on how to chose between the two formulations of sparse multinomial logistic regression models, MultinomSparseSym and MultinomSparseRef. If a high level of homogeneity exists among vectors $\bbeta^*_1, \ldots, \bbeta^*_K$ (or $\bdelta^*_1, \ldots, \bdelta^*_{K-1})$, then estimation efficiency is expected to be much higher when working with MultinomSparseSym (or equivalently MultinomDataSharedRef).  

The estimation of several parameter vectors considered here is closely related to multi-task learning \cite{Evgeniou}, for which a number of other structured sparsity inducing norms have been proposed in the literature, including the group lasso and generalized fused lasso \cite{LouniciTsyb,GenFused}.  However, we shall mention that the group lasso is not well suited when the identification of heterogeneities is of primary interest. On the other hand, the generalized fused lasso has shown good properties in the context of stratified regression models, both under generalized linear models \cite{GenFused}, survival models \cite{ReulenKneib} and binary graphical models \cite{NadimBinGraph}. Its extension to conditional logistic regression models or multinomial logistic models will be the focus of future work.

\section*{Acknowledgements}

This work was partially supported by the French National Cancer Institute (L'Institut National du Cancer; INCA) (grant number 2015-166; PI: S. Rinaldi). The authors are grateful to the Principal Investigators of each of the EPIC centres for sharing the data we used in our illustrative example.

 \bibliographystyle{apalike}

\appendix

\section{Additional technical details}

\subsection{Proof of \eqref{eq:medianL1sym}}\label{proof:medianL1sym}

For any $\bnu\in\R^p$, maximizers $\hbbeta_1, \ldots, \hbbeta_K$ of the criterion penalized in the glmnet package are such that:
\begin{align*}
\cL(\hbbeta_1, \ldots, \hbbeta_K) - \lambda \sum_{k=1}^K \|\hbbeta_k\|_1 & \geq  \cL(\hbbeta_1-\bnu, \ldots, \hbbeta_K-\bnu) - \lambda \sum_{k=1}^K \|\hbbeta_k -\bnu\|_1\\
&\geq  \cL(\hbbeta_1, \ldots, \hbbeta_K) - \lambda \sum_{k=1}^K \|\hbbeta_k -\bnu\|_1. 
\end{align*}
 Therefore,  $\sum_{k=1}^K \|\hbbeta_k\|_1 \leq \sum_{k=1}^K \|\hbbeta_k -\bnu\|_1$ for all $\bnu\in\R^p$ which establishes \eqref{eq:medianL1sym}.

\subsection{Equivalence between MultinomDataSharedRef and MultinomSparseSym}

With the particular choice $\tau_k=1$ for all $k\in[K-1]$, MultinomSparseSym consists in maximizing the criterion $\LL(\bbeta_1, \ldots, \bbeta_K) - \lambda \sum_{k=1}^K  \|\bbeta_k\|_1$. For any given $(\bbeta_1, \ldots, \bbeta_K)$, set $\bmu = -\bbeta_K$ and $\bgamma_k = \bbeta_k$ for all $k\in[K-1]$. Then we have
\begin{align*}
&\LL(\bbeta_1, \ldots, \bbeta_K) - \lambda \sum_{k=1}^K  \|\bbeta_k\|_1 \\
&\quad\quad = \LL(\bgamma_1, \ldots, \bgamma_{K-1}, -\bmu) - \lambda( \| - \bmu\|_1 +  \sum_{k=1}^{K-1}  \|\bgamma_k\|_1)\\
&\quad\quad = \LL(\bmu+\bgamma_1, \ldots, \bmu+\bgamma_{K-1}, {\bf 0}_p) - \lambda (\|\bmu\|_1 + \sum_{k= 1}^{K-1} \| \bgamma_k\|_1)\\
&\quad\quad=  L(\bmu + \bgamma_1, \ldots,\bmu + \bgamma_{K-1})  - \lambda (\|\bmu\|_1 + \sum_{k=1}^{K-1} \|\bgamma_k\|_1),
\end{align*}
which is exactly the criterion maximized by MultinomDataSharedRef for the the particular choice $\tau_k=1$ for all $k\in[K-1]$.

\subsection{Matrix formulation of the log-likelihood \eqref{eq:logVraisPolyt}}

Denote the indicator function by $\1(\cdot)$. For $k\in[K-1]$, introduce ${\cal Y}^{(k)}\in\R^n$ with ${\cal Y}^{(k)}_i = \1(Y_i = k)$, for all $ i\in[n]$. Further introduce the vector of binary variables ${\cal Y}\in\R^{n(K-1)}$ and the matrix $\bcX\in \R^{n(K-1)\times (K-1)p}$ defined as  
$$
{\cal Y} = \left(
\begin{array}{c}
{\cal Y}^{(1)} \\
\vdots\\
{\cal Y}^{(K-1)}
\end{array}
\right)
\quad{\rm and}\quad
\bcX = \left( 
\begin{array}{c c c }
\bX& \hdots &{\bf 0}_{n,p}  \\
 \vdots & \ddots & \vdots \\
{\bf 0}_{n,p}&\hdots  &  \bX  
\end{array}
\right),
$$ 
where $\bX$ is the $n\times p$ matrix containing the $n$ observations $(\bx_i)_{1\leq i\leq n}$ of the $p$ predictors. Finally set ${\boldsymbol 1}_n$ the vector of length $n$ whose components are all equal to 1, and ${\bf J} = ({\bf I}_n, \ldots, {\bf I}_n)$ the $n\times n(K-1)$ matrix whose each of the $(K-1)$ blocks is the identity matrix of order $n$, ${\bf I}_n$. Then, setting  $\bdelta=(\bdelta_1^T, \ldots, \bdelta_{K-1}^T)^T$, the log-likelihood (\ref{eq:logVraisPolyt}) can be rewritten more compactly as
\begin{align}
L(\bdelta_1, \ldots, \bdelta_{K-1})  &= {\cal Y}^T \bcX \bdelta - \left\{\log\left( {\boldsymbol 1}_n + {\bf J} \exp(\bcX\bdelta)\right) \right\}^T{\boldsymbol 1}_n.  \label{eq:logVraisPolyt2}
\end{align}

\end{document}